\documentclass[5p,10pt]{elsarticle}
\makeatletter
\let\old@ssect\@ssect 
\makeatother
\usepackage{mathtools}
\usepackage{amssymb}
\usepackage{enumerate}
\usepackage{definice}
\usepackage{color}
\usepackage{diagbox,multirow}
\usepackage{multirow}
\usepackage{float}
\usepackage{comment}
\usepackage{amsmath,amsfonts}
\usepackage{algorithmic}
\usepackage{algorithm}
\usepackage{textcomp}
\usepackage{stfloats}
\usepackage{graphicx}
\usepackage{xfrac}
\usepackage{tikz}
\usepackage{bigfoot}
\usepackage{graphicx}      

\usepackage{enumitem}

\newcommand{\MATLAB}{\textsc{MATLAB}}

\def\l{^{L}}

\newcommand{\true}{(\raisebox{0.67mm}{\includegraphics[scale=0.6]{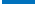}})}
\newcommand{\est}{(\raisebox{0.67mm}{\includegraphics[scale=0.6]{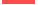}})}
\newcommand{\error}{(\raisebox{0.67mm}{\includegraphics[scale=0.6]{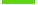}})}

\newcommand{\tightoverset}[2]{%
  \mathop{#2}\limits^{\vbox to -.5ex{\kern-0.75ex\hbox{$#1$}\vss}}}
  
\begin{document}
\begin{frontmatter}

\title{\hspace{-0mm}Time-Varying Kernel Density Deconvolution using Random Finite Set Theory for State Noise Identification}
\author[First]{Jind\v{r}ich Dun\'{i}k, Oliver Kost, Jan Krej\v{c}\'{i}, and Ond\v{r}ej Straka} 

\address[First]{Department of Cybernetics\\Faculty of Applied
Sciences, University of West Bohemia in Pilsen, \\Univerzitn\'{i} 8, 306 14 Pilsen, Czech Republic \\E-mails: \{dunikj,kost,jkrejci,straka30\}@fav.zcu.cz}

\begin{abstract} 
This paper focuses on the noise density identification for state-space models of linear time-varying systems. The density knowledge is essential for the optimal design of any modern state estimator, fault detector, or controller. The proposed method builds on \textit{newly derived} algorithms for (\textit{i}) calculation of the measurement prediction residue being independent of the state and linear in state and measurement noises, and (\textit{ii}) kernel density deconvolution for process with non-stationary statistical properties based on the random finite set theory. The proposed method provides \textit{consistent} estimate of the state noise density \textit{without any prior knowledge} on distribution family, which is a unique property in noise identification state-of-the-art. The method is further enhanced by an automatic selection of the user-defined parameters, and is evaluated in numerical examples. The examples demonstrate not only the superior performance of the proposed method in terms of accuracy and robustness among others in ultra-wide-band-inspired tracking scenario, but also the importance of the correct state noise density identification for optimal state estimation and fault detection. The paper is supplemented with a \MATLAB\textregistered\ implementation.
\end{abstract}

\begin{keyword}
Linear systems, Noise identification, Probability density function, Kernel density, Density deconvolution, Measurement difference method.
\end{keyword}

\end{frontmatter}

\section{Introduction}
Modern control, fault detection, and signal-processing algorithms rely on the availability of an accurate model of the system, often in the state-space form \citep{Si:06,CrJu:12}.  The model describes the time behaviour of the unknown (but inferred) system quantities called the state and the relation between the state and the available noisy measurement produced by sensors. The state-space model can virtually be seen as a composition of two sub-models: \textit{deterministic} and \textit{stochastic}. The former captures system behaviour arising often from the first principles based on physical, kinematical, chemical, mathematical, biological, or other laws and rules. The latter characterises the uncertain behaviour of the state and measurement, where the uncertainties are modelled by noises described in the stochastic framework. 

Due to the inherent difficulty in uncertainty modelling using first principles, noise characteristics are at least partially identified from data.
Since the seventies, thus, noise identification has attracted significant attention, and a wide range of methods for noise \textit{moment} and \textit{density parameter} estimation have been proposed.

\subsection{Estimation of Moments}
Most methods have predominantly focused on the estimation of the state and measurement noise covariance matrices. 
These methods can be classified into five categories; \textit{(i)}~correlation \citep{Meh:70,Be:74,OdRaRa:06,DuKoSt:18}, \textit{(ii)}~Bayesian \citep{SaNu:09,ArOzOrGu:15,MaLaWaZeWaPaMo:18,ChChZhZh:21}, \textit{(iii)}~maximum likelihood \citep{ShSt:00}, \textit{(iv)}~covariance matching \citep{SaHu:69}, and \textit{(v)}~machine-learning \citep{CoIt:24}. Besides the covariance matrix estimation, some of the correlation methods provide estimates of the means and higher-order moments (such as skewness and kurtosis) \citep{AkJoJo:07,KoDuSt:22}.

\subsection{Estimation of Density Parameters}
While knowing the noise moments is adequate for certain estimation methods, such as the (extended, unscented etc.) Kalman filter \citep{Si:06}, other techniques for state estimation (such as particle or grid-based filters \citep{Be:99,BaLiKi:01}) or fault detection necessitate more comprehensive description of both the state and measurement noises. In particular, they require the probability density functions (PDFs) of these noises. Therefore, several methods have also been designed to characterise the noise PDFs. 

In this context, there are methods assessing the Gaussianity of the state and measurement noises \citep{DuKoStBl:20}. If the noises are Gaussian, a wide range of efficient signal processing or control algorithms can be used. In addition to assessing noise PDFs, there are methods that directly estimate the unknown parameters of these noise PDFs assumed to belong to a \textit{known} family, such as Student's t-distribution with unknown degrees of freedom or Gaussian sum PDF with unknown moments and weights of the individual modes. Here, we can differentiate between \emph{direct} and \emph{indirect} methods. The former methods estimate the parameters of a noise PDF directly, such as by augmenting the state vector and assuming the analytically tractable form for the posterior \citep{SaNu:09}. The latter methods calculate a set of noise moments and use these moments to determine the PDF parameters through the method of moments \citep{Ka:93,KoDuSt:22}.  

\subsection{Research Gap}
The direct and indirect methods for noise PDF parameter estimation are, however, applicable only under the assumption of the \textit{known} noise PDF family. The literature lacks a method for estimating the noise PDF \emph{without} this a priori knowledge, which is the ultimate objective of this paper. 


\subsection{Contribution}
In this paper, we address this limitation and propose the method for non-parametric estimation of a noise PDF \textit{without} any prior knowledge on the distribution family. The stress is laid on state noise PDF estimation for linear time-varying models with full-column rank measurement matrix, widely encountered in tracking and navigation.  The proposed method is based on \textit{newly} developed fundamental algorithms for:
\begin{itemize}
    \item[(\textit{i})] Residuum calculation being a linear function of the state and measurement noise, 
    \item[(\textit{ii})] Kernel density deconvolution method for residuum with time-varying properties derived using random finite sets,
\end{itemize}
and their \textit{combination} in a unified framework that yields method's robust convergence and consistency guarantees. The performance of the proposed method is evaluated in numerical simulations (inspired by ultra-wideband (UWB) tracking scenario) using the publicly available codes\footnote{\small{https://github.com/IDM-UWB/StateNoisePDFIdentification}} and compared with the state-of-the-art techniques for noise identification under assumption of known  PDF family.
\newline




The rest of the paper is organised as follows. In Section~2, the state-space model is defined and possible application areas are particularised. Section 3 then focuses on the development of the proposed state noise PDF identification method and Section 4 addresses implementation aspects and convergence of the method. The algorithm summary and parameter selection, and numerical validation are provided in Sections 5 and 6, respectively. Concluding remarks are drawn in Section 7.


\section{System Description and Formulation}
In this paper, we consider the linear time-varying (LTV) discrete-in-time stochastic dynamic model in state-space representation with additive noises
\begin{subequations}\label{model}
	\begin{align}
		\bfx_{k+1}&=\bfF_k\bfx_k+\bfG_k\bfu_k+\bfw_k,\label{eqSt}\\
		\bfz_{k}&=\bfH_k\bfx_k+\bfv_k,\label{eqMe}
	\end{align}
\end{subequations}
where $ k=0,1,2,\ldots,\tau $ denotes a time step and the vectors $\bfx_k\in\real^{n_x}$, $\bfu_k\in\real^{n_{u}}$, and $\bfz_k\in\real^{n_{z}}$ represent the immeasurable state of the system, the available control, and the available measurement at time instant $k$, respectively. The matrices $\bfF_k\in\real^{n_x\times n_x}$, $\bfG_k\in\real^{n_x\times n_{u}}$, and $\bfH_k\in\real^{n_{z}\times n_x}$ are \textit{known} state, control, and measurement matrices, respectively. The \text{measurement matrix} is supposed to have \textit{full} {column rank}, i.e., the inverse of the matrix $(\bfH_k^T\bfH_k)$ exists, $\forall k$, which implies that $n_z\geq n_x$. 

The state and measurement noises  $\bfw_k\in\real^{n_x}$  and $ \bfv_k\in\real^{n_z}$, respectively,  are unknown with 
    the \textit{measurement noise} PDF $p_{\bfv_k}(\bfv_k)$ assumed to be known, 
    and the \textit{state noise} PDF $p_\bfw(\bfw_k)$ assumed to be  \textit{unknown} (and identified).

\subsection{Motivation and Applications}
The considered state-space model structure, namely Eq.~\eqref{model} with the assumption $n_z\geq n_x$, is widely used in many areas, including the GNSS positioning \citep{Gr:08,GuVe:88,MaLaWaZeWaPaMo:18,ChChZhZh:21}, multi-radar and UWB tracking \citep{BaLiKi:01,NuArPiGu:15a,YaPaXuLu:22}, economics \citep{HaKo:09,ShSt:00}, or chemical processes \citep{Br:86}.


Taking the global navigation system (GNSS) \textit{positioning} as an example, the state $\bfx_k$ consists of the unknown estimated vehicle's position and velocity. The measurement vector contains available pseudo-ranges and delta-ranges between the satellites and vehicle receiver. Considering the four available GNSS constellations, the number of available measurements $n_z$ anywhere on Earth is much higher than the state dimension $n_x$.

In this case, the vehicle dynamics can be described by a variety of user-selected models (i.e., state equations) such as the nearly constant velocity, coordinated turn-rate, or Singer models~\citep{BaLiKi:01}. The relation of the state to the sensed data is given by first principles, leading to the measurement equation~\citep{Gr:08}. Then: 
\begin{itemize}
    \item The measurement noise $\bfv_k$ is predominantly determined by the sensor noise, which can be well described and modelled using data-sheets or calibration procedures based on the available set of measurement $\bfz_k, \forall k,$ \citep{Ro:03,Gr:08,NiFuMeSi:16,KoDuStDa:23}, and thus, its description $p_\mathbf{v_k}(\mathbf{v}_k)$ can be treated as \emph{known}.    
    \item The state (or process) noise $\bfw_k$ represents the discrepancy between (unknown) true and modelled dynamics. This means that specification of the  matrices $\bfF_k, \bfG_k$ affects the noise characteristics. Consequently, the state noise PDF \textit{cannot} be determined from the first principles, but its PDF \( p_\mathbf{w}(\mathbf{w}_k) \) must be \emph{identified}.
\end{itemize}

\subsection{Goal of the Paper}

The \textit{goal} of the paper is to 
develop a method for the identification of the state noise PDF $p_\bfw(\bfw_k)$ without \textit{any} prior assumption on its PDF family.
The identification is based on the measured and input data $\bfz^k=[\bfz_0,\bfz_1,\ldots,\bfz_k]$ and $\bfu^{k-1}=[\bfu_0,\bfu_1,\ldots,\bfu_{k-1}]$, known model matrices $\bfF_k$, $\bfG_k$, $\bfH_k$, and known measurement noise PDF $p_{\bfv_k}(\bfv_k), \forall k$. 

The novel method for the \textit{state noise PDF identification based on the non-stationary kernel density deconvolution}, further abbreviated as the KDI-SNI, is proposed in the following three sections. The method properties are analysed, and algorithm parametrisation and implementation notes are provided. 

\section{State Noise PDF Identification using Kernel Deconvolution via Random Finite Sets} \label{sec:noise_covariance_identification}
For the proposed KDD-SNI method, two key components are newly developed for the proposed method: \emph{(i)} residue calculation, which is independent of the state and linear w.r.t. the state and measurement noises, \emph{(ii)} kernel density deconvolution (KDD) for non-stationary processes using random finite sets. 

Developed components and their combination \textit{enables} derivation of a novel approach for non-parametric state noise density identification. 

\subsection{Residue Design and its Calculation}
The proposed method is based on the calculation of the residue (or measurement prediction error), which is a \textit{linear} function of the state and measurement noises and is independent of the state. The residue, denoted with $\bfr^\bfz_k$, is defined as the \textit{measurement} prediction error, i.e., as a difference between the measurement $\bfz_k$ and its \textit{one-step} prediction $\hbfz_k$ \citep{DuKoSt:18,KoDuSt:22}, and it reads
\begin{align}
    \bfr^\bfz_k \triangleq \bfz_k-\hbfz_k \,, \label{eq:resz}
\end{align}
with 
\begin{align}
    \hbfz_k=\bfH_k\bfF_{k-1}\bfH^\dagger_{k-1}\bfz_{k-1}+\bfH_k\bfG_{k-1}\bfu_{k-1}\,,\label{eq:measPred}
\end{align}
where $\bfH^\dagger_{k-1}=(\bfH_{k-1}^T\bfH_{k-1})^{-1}\bfH_{k-1}^T$ stands for the left pseudoinverse. 
The prediction $\hbfz_k$ is \textit{not} optimal in terms of the mean-squared-error; however, the resulting residual \eqref{eq:resz} can be shown, w.r.t. \eqref{eqSt}, to be a linear function of the state and measurement noises as
\begin{align}
    \bfr^\bfz_k&=\bfH_k\bfx_k+\bfv_k-\bfH_k\bfF_{k-1}\bfH^\dagger_{k-1}(\bfH_{k-1}\bfx_{k-1}+\bfv_{k-1})\nonumber\\
    &=\bfv_k+\bfH_k\bfw_{k-1}-\bfH_k\bfF_{k-1}\bfH^\dagger_{k-1}\bfv_{k-1}.\label{eq:measPred2}
\end{align}

To enable state noise PDF $p_\bfw(\bfw_k)$ estimation, we construct the \textit{state-noise-related} residue, denoted with $\bfr_k$, using \eqref{eq:resz} and \eqref{eq:measPred2} as
\begin{subequations}\label{eq:resx_all}
\begin{align}
    \bfr_k& \triangleq \bfH_k^\dagger\bfr^\bfz_k = \bfH_k^\dagger (\bfz_k-\hbfz_k) \label{eq:resxComp}\\
    &=\bfw_{k-1}+\bfH_k^\dagger\bfv_k-\bfF_{k-1}\bfH^\dagger_{k-1}\bfv_{k-1}.\label{eq:resx}
\end{align}
\end{subequations}
While Eq. \eqref{eq:resxComp} shows how the state-noise-related residue is calculated from actual measurements $\bfz_k$ and $\bfz_{k-1}$, Eq.~\eqref{eq:resx} reveals that $\bfr_k$ is a sum of
\begin{itemize}
    \item State noise $\bfw_{k-1}$ and 
    \item Linear transformation of the measurement noise $\bfv_k$ and $\bfv_{k-1}$,
\end{itemize}
which is the \textit{key} aspect for the following derivation of the proposed method.

Since the measurement noise PDF $p_{\bfv_k}(\bfv_k)$ and model matrices $\bfF_k, \bfH_k, \forall k$ are \textit{known}, we can rewrite \eqref{eq:resx} as 
\begin{align}
    \bfr_k=\bfw_{k-1}+\bfnu_k,\label{eq:resx_sum}
\end{align}
where the \textit{transformed measurement noise} $\bfnu_k=\bfH_k^\dagger\bfv_k-\bfF_{k-1}\bfH^\dagger_{k-1}\bfv_{k-1}$ has \textit{known} time-varying density $p_{\bfnu_k}$.


\subsection{State Noise Identification by Time-Varying Density Deconvolution via Random Finite Sets}
Let us assume for the moment that the density  ${p}_{\bfr_k}$ of the residue $\bfr_k$ \eqref{eq:resx_sum} is \textit{known}. Then, taking into account the \textit{known} transformed measurement noise PDF $p_{\bfnu_k}$ in \eqref{eq:resx_sum}, we can, in principle, estimate the state noise PDF $p_\bfw$ using the KDD \citep{StCa:90,BrMo:12,Che:17}. The KDD is a statistical method used to estimate the PDF of a latent variable, in our case the state noise PDF $p_\bfw$, given the observed (and known) data $\bfr_k$ \eqref{eq:resxComp}, which is ``contaminated'' by the noise, in our case $\bfnu_k$ \eqref{eq:resx_sum}. 

Unfortunately, the state-of-the-art KDDs are designed under assumption of time-invariant description of involved quantities \citep{Che:17}. In our case, however, the transformed measurement noise PDF $p_{\bfnu_k}(\bfnu_k)$, and thus, the density ${p}_{\bfr_k}$, varies in time due to LTV model \eqref{model}. 
To address this limitation, we design a novel \textit{convolution equation} for {single-pass KDD} for \textit{time-varying} densities of the residuum $\bfr_k$ and the noise $\bfnu_k$.

\subsubsection{Derivation of Convolution Equation for Time-Varying Densities}
First, define a subset of residue samples, where each element is \textit{independent}. Due to its construction \eqref{eq:resx}, the residue $\bfr_k$ is correlated with the preceding $\bfr_{k-1}$ only, because of the sum of measurement noise realisations in two subsequent time instants. Therefore, a subset of independent residues can simply be defined as $R_{s}=\{ \bfr_k\}_{k\in\calK}$ with $\calK=\{0,2,4,\ldots,\tau\}$ ($\tau$ even). The cardinality of the subset $\calK$ is further denoted as $\kappa$.

Second, notice that 
each residuum sample $\bfr_k$~\eqref{eq:resxComp} with $k\in\calK$ can be viewed as a realization of the \textit{random finite set} (RFS) $R_{s}$, where the residuum exists with probability one and is generated according to the distribution $p_{\bfr_k}$. Following this interpretation, the RFS given by $R_k=\{\bfr_k\}$ has the PDF
\begin{align}
    p_{R_k}(R_k) = \begin{cases}
        p_{\bfr_k}(\bfr_k), & \text{if } R_k=\{\bfr_k\} \,, \\
        0, & \text{otherwise} .
    \end{cases}\label{eq:multiber} 
\end{align}
Following~\citep{Mahler-Book:2014}, we note the PDF $p(\cdot)$ of any RFS is normalised such that its \emph{set integral} defined by
\begin{align}
     \int p_{}(R) \delta R = \sum_{i=0}^{+\infty} \frac{1}{i!} \int p_{}(\{\bfr^1,\dots,\bfr^i\}) d\bfr^1\cdots d\bfr^i \,,
\end{align}
is equal to one, where $R$ and $\bfr^j$ are integration variables.

Since the residue samples $\bfr_k$~\eqref{eq:resxComp} are mutually distinct and independent, the union $R_{s}=\cup_{k\in\calK} R_k = \{\bfr_k\}_{k\in\calK}$ 
has \emph{multi-Bernoulli} distribution. 
The PDF expression of a multi-Bernoulli-distributed RFS can be found in~\citep[Sec.~3.4.3]{Mahler-Book:2014}. However, for the purpose of the time-varying KDD design, the first moment of $R_s$ is the only information needed.

The first statistical moment $D(\cdot)$ of any RFS, also known as the \emph{probability hypothesis density} (PHD) or the \emph{intensity}, is a function of one variable $\mathbf{r}$, for which
\begin{align}
    D_{}(\bfr) = \int p_{}(R \cup \{\bfr\}) \delta R \, . \label{eq:PHD-def}
\end{align}
The PHD $D_{}(\bfr)$~\eqref{eq:PHD-def} is an un-normalized density function whose ordinary integral $\int_B D_{}(\bfr) d\bfr$ is the expected cardinality of the RFS intersecting the region~$B$.

The PHD of each Bernoulli-distributed $R_k$ is equal to $D_{R_k}(\bfr_k)=p_{\bfr_k}(\bfr_k)$.
As the PHD of a union of independent random finite sets is the sum of PHDs of the united sets~\citep[p.~94]{Mahler-Book:2014}, the PHD of $R_{s}=\cup_{k\in\calK} R_k$ is given by
\begin{align}
    D_{R_{s}}(\bfr) = \sum_{k\in\calK} D_{R_k}(\bfr) = \sum_{k\in\calK} p_{\bfr_k}(\bfr) \,,
\end{align}
where $\bfr$ is a free variable.
Using~\eqref{eq:resx_sum}, the density $p_{\bfr_k}(\cdot)$ is given by the convolution of $p_{\bfw}(\cdot)$ and $p_{\bfnu_k}(\cdot)$.
Therefore,
\begin{align}
    D_{R_{s}}(\bfr) = \sum_{k\in\calK} \int p_{\bfw}(\bfr-\bfnu_k)p_{\bfnu_k}(\bfnu_k)d\bfnu_k \, . \label{eq:PHD-derivation-R}
\end{align}
Normalizing~\eqref{eq:PHD-derivation-R} by the expected cardinality $\kappa = \int D_R(\bfr) d\bfr$ of $R$ in the entire space and taking the summation (i.e., averaging over $k\in\calK$) into the convolution yields the desired final relation for the convolution
\begin{align}
    \frac{1}{\kappa}\sum_{k\in\calK}{p}_{\bfr_k}(\bfr)&=\int p_{\bfw}(\bfr-\bfnu_k)\frac{1}{\kappa}\sum_{k\in \calK}p_{\bfnu_k}(\bfnu_k)d\bfnu_k \,,
    \label{eq:conv_ltv}
\end{align}
meaning also that the single-time instant convolution is
\begin{align}
    {p}_{\bfr_k}(\bfr_k)&=\int p_{\bfw}(\bfr_k-\bfnu_k)p_{\bfnu_k}(\bfnu_k)d\bfnu_k
    \label{eq:conv_ltv_single} \,.
\end{align}

\subsubsection{Characteristic Function Definition and Deconvolution}
Due to \eqref{eq:resx_sum} and the independence of the random variables $\bfw$ and $\bfnu_k$ for $k\in\calK$, i.e. $p_{\bfw,\bfnu_k}(\bfw,\bfnu_k)=p_{\bfw}(\bfw)\,p_{\bfnu_k}(\bfnu_k)$, the PDF of $p_{\bfr_k}(\bfr_k)$ \eqref{eq:conv_ltv_single} is the convolution of the PDFs of the latent and noise variables $p_{\bfw}(\bfw)$ and $p_{\bfnu_k}(\bfnu_k)$. The convolution can be rewritten into the frequency domain as
\begin{align}\label{eq:CFproduct}
{\calF}_{\bfr_k}(\bft)=\calF_{\bfw}(\bft)\calF_{\bfnu_k}(\bft),
\end{align}
where the notation
\begin{align}
 \calF_{\bfx}(\bft)=\mean[e^{i\bft\bfx}]=\int p_{\bfx}(\bfo)e^{i\bft\bfo}d\bfo   
\end{align}
stands for the \textit{characteristic function} (CF) of a random variable $\bfx$ obtained by the Fourier transformation (FT) of its PDF $p_{\bfx}(\bfx)$, with $\bft$ acting as a \textit{frequency}-like (real) vector variable, $i=\sqrt{-1}$, and $\mean[\cdot]$ is the expectation.

Appling \eqref{eq:CFproduct} and using the linearity of the FT, we can rewrite \eqref{eq:conv_ltv} using the sum of the CFs as
\begin{align}
    \sum_{k\in\calK}{\calF}_{\bfr_k}(\bft)={\calF}_{\bfw}(\bft)\sum_{k\in\calK}\calF_{\bfnu_k}(\bft).\label{eq:Fw_ltv}
\end{align}
Then, the state noise PDF $\hat{p}_\bfw(\bfw)$ can be computed by the inverse Fourier transform (iFT) of the state noise CF given by
\begin{align}
    {\calF}_{\bfw} (\bft)
    &=\frac{\sum_{k\in\calK}{\calF}_{\bfr_k} (\bft)}{\sum_{k\in\calK}\calF_{\bfnu_k} (\bft)}.\label{eq:Fwestim_ltv_full_concept}
\end{align}
Note that this state noise PDF $\hat{p}_\bfw(\bfw)$ can be seen as fused density in the sense of the \textit{arithmetic mean} over all time\discretionary{-}{-}{-}varying residue and noise PDFs for all $ k\in\calK$, which is proven in Appendix.

\subsection{Residue Kernel Densities Construction}
Although \eqref{eq:Fwestim_ltv_full_concept} gives us a relation for state noise PDF calculation, we cannot directly use it as the true PDF $ {p}_{\bfr_k}(\bfr)$ (and thus the CF ${\calF}_{\bfr_k} (\bft)$) is \textit{unknown}. However, we can construct an estimate of the residue PDF using the available samples of the residue $\bfr_k,\ k\in\calK$, \eqref{eq:resxComp} using the \textit{kernel density} (KD) approximation \citep{Che:17}. Then, the \textit{single-sample} KD is defined as
\begin{align}
    \hat{p}_{\bfr_k}(\bfr)=|\bfB|^{-1/2}K\left(\bfB^{-1/2}(\bfr-\bfr_k)\right), \ \forall k,
    \label{eq:kernel_r_LTV}
\end{align}
where $\bfB\in\real^{n_x\times n_x}$ represents the smoothing (symmetric and positive definite) matrix whose determinant is denoted with $|\bfB|$, and $K(\cdot)$ is the \emph{kernel function} in the form of a multivariate probability density that is symmetric about the origin.

\subsection{State Noise Density Identification using Kernel Density Deconvolution}
Replacing the unknown true residue CFs ${\calF}_{\bfr_k}(\bft), \forall k,$ in \eqref{eq:Fwestim_ltv_full_concept} with their known KD-based counterparts $\hat{\calF}_{\bfr_k}(\bft), \forall k,$ calculated by the FT of $\hat{p}_{\bfr_k}(\bfr)$ \eqref{eq:kernel_r_LTV}, the state noise PDF estimate $\hat{p}_\bfw(\bfw)$ can be computed by the iFT of the state noise CF estimate given by
\begin{subequations}\label{eq:Fwestim_ltv_full_all}
    \begin{align}
        \!\!\!\hat{\calF}_{\bfw} (\bft)
        &=\frac{\sum_{k\in\calK}{\hat{\calF}}_{\bfr_k} (\bft)}{\sum_{k\in\calK}\calF_{\bfnu_k} (\bft)}\label{eq:Fwestim_ltv_full_concept_est}\\
        &=\frac{\int|\bfB|^{-1/2}\sum_{k\in \calK}K\!\left(\bfB^{-1/2}(\bfr-\bfr_k)\right)e^{i\bft\bfr}d\bfr}{\int \sum_{k\in\calK} p_{\bfnu_k}(\bfnu) e^{i\bft\bfnu}d\bfnu}.\!\!\!\label{eq:Fwestim_ltv_full}
    \end{align}
\end{subequations}






\section{Regularised KDD and Estimate Consistency}
The KDD-SNI method for state noise identification was derived in the previous section with the emphasis mainly on the sound mathematical derivation, rather than on the stable implementation. In this section, we address some aspects of the KDD-SNI method implementation and the closely related estimate properties (in terms of the consistency).

\subsection{Regularized KDD with Improved Numerical Stability}
The introduced KDD density deconvolution is, in general, ill-posed problem in the sense of Hadamard \citep{Jo:09}, as it requires the inversion of the characteristic function of the transformed measurement noise $\bfnu_k$, which may be arbitrarily
close to zero at high frequencies. To overcome this instability, regularized versions of the deconvolution estimator have been proposed that stabilize the inversion by controlling noise amplification \cite{CoRoTa:06,Jo:09}. Below, we introduce a regularized version of the estimator for the time-varying KDD \eqref{eq:Fwestim_ltv_full_all}, which can be rewritten in the form 
\begin{align}
    \hat{\calF}_{\bfw} (\bft)
    &=\frac{\sum_{k\in\calK}\tfrac{1}{\kappa}{\hat\calF}_{\bfr_k} (\bft)}{\sum_{k\in\calK}\tfrac{1}{\kappa}\calF_{\bfnu_k} (\bft)}.\label{eq:Fwestim_ltv_full_concept2} 
\end{align}
Since the convex combination of the CF (i.e., their weighted sum with positive weights summed to one) in \eqref{eq:Fwestim_ltv_full_concept2} is the CF \citep{Lu:70}, the state noise CF estimator \eqref{eq:Fwestim_ltv_full_concept2} reads
\begin{align}
    \hat{\calF}_{\bfw} (\bft)
    &=\frac{{\hat\calF}_{\bfr,\kappa} (\bft)}{\calF_{\bfnu,\kappa} (\bft)}.\label{eq:Fwestim_ltv_full_concept3} 
\end{align}
where the notation $\calF_{\bfnu,\kappa}=\sum_{k\in\calK}\tfrac{1}{\kappa}\calF_{\bfnu_k} (\bft)$ was used.

Then, the regularized inversion in \eqref{eq:Fwestim_ltv_full_concept3} via the least squares leads to the \textit{regularized} estimator
\begin{align}
\hat{\calF}_{\bfw}^{\alpha}(\bft)
=
\frac{
\hat{\calF}_{\bfr,\kappa}(\bft)\,\overline{\calF_{\bfnu,\kappa}(\bft)}
}{
|\calF_{\bfnu,\kappa}(\bft)|^2
}\mathbf{1}\{|\calF_{\bfnu,\kappa}(\bft)/\ell_s(\bft)|^2\geq\alpha\},\label{eq:Fwestim_lti_full_concept_regularised}
\end{align}
where $\overline{\calF_{\bfnu,\kappa}(\bft)}$ is the complex conjugate of ${\calF_{\bfnu\kappa}(\bft)}$ and the regularisation function $\ell_s(\bft)$ and the threshold $\alpha$ are detailed in \citep{Jo:09}.

\subsection{Estimate Consistency}
To discuss the consistency of the estimate \eqref{eq:Fwestim_lti_full_concept_regularised}, we can adopt the results of \citep{Jo:09}, where it was proven that for the estimate $\hat{p}_\bfw^\alpha$ obtained by the iFT of $\hat{\calF}_{\bfw}^{\alpha}(\bft)$, it holds that
\begin{align}
    \mean||\hat{p}_\bfw^\alpha(\bfw_k)-{p}_\bfw(\bfw_k)||_s^2=O(1)
\end{align}
provided that $p_\bfw$ belongs to the Sobolev space $H_s$, $s\geq0$. That is, $\hat{p}_{\bfr,\kappa}^\alpha$ is consistent estimate of ${p}_{\bfr,\kappa}$ meaning $\mean||\hat{p}_{\bfr,\kappa}(\bfr_k)-{p}_{\bfr,\kappa}(\bfr_k)||^2=O(1)$ as $\kappa\rightarrow\infty$ \citep{Che:17}, and suitably selected regularisation function $\ell_s(\bft)$ defined in \citep{Jo:09}. In other words, it means that the estimate error $\hat{p}_\bfw^\alpha(\bfw_k)-{p}_\bfw(\bfw_k)$ tends to zero as the sample size $\kappa$ increases for a wide class of densities including Gaussian, Laplace,  Rayleigh, Student's-t, or Beta densities to name a few.



\section{Algorithm and Density Parametrisation}
\subsection{Algorithm}\label{sec:alg}
Below is the pseudo-code for the proposed method for identifying the state noise PDF. The pseudo-code is consistent with the \MATLAB\textregistered\ implementation\footnotemark[\value{footnote}] of the numerical examples given in the subsequent section. 
\newline
\rule{8.8cm}{1pt}
\newline
\textbf{Algorithm}: State Noise PDF Identification by KDD-SNI
\newline
\vspace*{-4mm}
\newline
\rule{8.8cm}{0.5pt}
	\begin{enumerate}[label=(\roman*)]
        \item Specify the model matrices $\bfF_k, \bfG_k, \bfH_k$, for $k=0,1,$ $\ldots,\tau$, the measurement noise PDF $p_{\bfv_k}(\bfv_k)$.
		\item Obtain measurement data $\bfz_k$ and control $ \bfu_k, \forall k$. 
        \item Calculate the measurement prediction $\hbfz_k$ and the residue $\bfr_k,\forall k$, according to \eqref{eq:resxComp} with \eqref{eq:resz} and \eqref{eq:measPred}.
        \item Define the set $\calK$, KD family, and parameters. Based on the residues $\bfr_k, k\in\calK$, construct KDs $\hat{p}_{\bfr_k}$ \eqref{eq:kernel_r_LTV}.
        \item Based on the model matrices and measurement noise PDF $p_{\bfv_k}$, determine the PDF of the transformed measurement noise $p_{\bfnu_k},\ \forall k,$. 
        \item Construct equidistant grid of points $\bfXi=\left[\bfxi^{(1)},\bfxi^{(2)},\right.$ $\left.\ldots,\bfxi^{(N)}\right]$ well-covering the support of the KDs $\hat{p}_{\bfr_k}$.
        \item Evaluate residue KD and transformed noise PDF at each grid point, i.e., calculate 
        \begin{subequations}\label{eq:PrPv_ltv}
            \begin{align}
                \hat{\calP}_{\bfr}&=\left[\sum_{k\in\calK}\hat{p}_{\bfr_k}(\bfxi^{(1)}),\ldots,\sum_{k\in\calK}\hat{p}_{\bfr_k}(\bfxi^{(N)})\right],\label{eq:Pr_ltv}\\
                \calP_{\bfnu}&=\left[\sum_{k\in\calK}p_{\bfnu_k}(\bfxi^{(1)}),\ldots,\sum_{k\in\calK}p_{\bfnu_k}(\bfxi^{(N)})\right].\label{eq:Pv_ltv}
            \end{align}            
        \end{subequations}
        \item Calculate CFs corresponding to the PDFs \eqref{eq:PrPv_ltv} 
        \begin{subequations}
            \begin{align}
                \hat{\calF}_{\bfr}&=\mathrm{FT}(\hat{\calP}_{\bfr}),\\
                \calF_{\bfnu}&=\mathrm{FT}(\calP_{\bfnu}).
            \end{align}
        \end{subequations}
        \item Estimate CF of the state noise $\bfw$ according to \eqref{eq:Fwestim_lti_full_concept_regularised}~as
        \begin{align}
              \hat{\calF}_{\bfw}=\hat{\calF}_{\bfr}\overline{\calF_{\bfnu}}\oslash|\calF_{\bfnu}|^2.\label{eq:Fw:num}
        \end{align}
        where $\oslash$ stands for the Hadamard (point-wise) division respecting the threshold $\alpha$ defined in \eqref{eq:Fwestim_lti_full_concept_regularised}.
        \item Calculate the state noise PDF estimate evaluated at the grid $\bfXi$ by the inverse Fourier transform
        \begin{align}
            \hat{\calP}_{\bfw}&=\mathrm{iFT}(\hat{\calF}_{\bfw}), \label{eq:pwEstim}
        \end{align}
        of which the \textit{real} part is shifted and normalised.
    \end{enumerate}
\rule{8.8cm}{1pt}
\newline

The FT and iFT can be calculated by the fast discrete algorithms. In \MATLAB\textregistered, these algorithms are available for arbitrary-dimensional arrays (e.g., routines \verb|fftn, ifftn|) as well as the algorithms for multidimensional KD design (e.g., routine \verb|mvksdensity|). Note that advanced density deconvolution algorithms are implemented in packages for \MATLAB\textregistered{} or for \textsc{R} \citep{CoRoTa:06,WaWa:11}.

\subsection{Sampling and Moment Calculation}
The non-parametric state noise PDF estimate $\hat{\calP}_{\bfw}$~\eqref{eq:pwEstim} can be conveniently written in a histogram-like point-mass density (PMD)
\begin{align}
    \hat{p}_{\bfw}(\bfw;\bfXi)\triangleq\sum_{i=1}^N\hat{\calP}_{\bfw}^{(i)}\mathbf{1}_{\Delta_{\bfxi}^{(i)}}(\bfw),\label{eq:pw_pmd}
\end{align}
where $\mathbf{1}_{\Delta_{\bfxi}^{(i)}}(\bfw)$ is an indicator function that equals one if a realisation of the state noise $\bfw$ is in the non-overlapping neighbourhood $\Delta_{\bfxi}^{(i)}$ of the $i$-th point $\bfxi^{(i)}_k$ and is zero otherwise.

The PMD can be easily (\textit{i}) used for sampling, (\textit{ii}) evaluated at arbitrary point, and (\textit{iii}) used for moment calculation \cite{MaDuSt:25}. For example, the mean and the covariance matrix of the state noise can simply be calculated as
\begin{align}
    \widehat\mean[\bfw]&= \sum_{i=1}^{N}\Delta_{\bfxi}^{(i)}\hat{\calP}_{\bfw}^{(i)}\bfxi^{(i)}\\
    \hbfQ(\bfB)&=\widehat\cov[\bfw]= \sum_{i=1}^{N}\Delta_{\bfxi}^{(i)}\hat{\calP}_{\bfw}^{(i)}\left(\bfxi^{(i)}-\widehat\mean[\bfw]\right)\left(\cdot\right)^T,\label{eq:pmd_cov}
\end{align}
where the explicit dependency of the estimate $\hbfQ (\bfB)$ on the user-defined kernel bandwidth $\bfB$ is highlighted. Note also that the form of the noise description \eqref{eq:pw_pmd} is a natural choice for the \textit{quantum-inspired} state estimation algorithms \cite{ChPaDuKrMaBr:26}.

\subsection{Parameters Selection via Moment Optimisation}
The KDD-SNI requires the specification of the kernel type, kernel parameters (such as the bandwidth $\bfB$) and possibly other regularisation parameters,  in the design of the KD $\hat{p}_{\bfr}(\bfr)$ in \eqref{eq:kernel_r_LTV}.

Without any prior information on the state noise PDF, the bandwidth can be determined adaptively or by using optimisation. The \textit{former} option involves the application of advanced adaptive deconvolution methods, where especially the kernel proprieties are determined automatically \citep{CoRoTa:06}. In this paper we propose to use the \textit{latter} option, taking advantage of the correlation methods, such as the measurement difference method (MDM) \citep{KoDuSt:22}. The MDM method provides \textit{unbiased} and \textit{consistent} estimate of the state noise covariance matrix $\hbfQ^\mathrm{MDM}$ independently of the true PDF $p_\bfw(\bfw_k)$. Then, the optimal selection of the kernel bandwidth (and potentially of other parameters) can be found by minimisation of the covariance-matrix-based metric \citep{FoMo:03} 
\begin{align}
    \bfB^*=\arg\min_{\bfB}d(\bfB),\label{eq:adapt_criterion}
\end{align}
where 
\begin{align}
d(\bfB)=\sqrt{\sum_{i=1}^{n_x}\ln^2\left(\lambda_i\left(\hbfQ(\bfB),\hbfQ^\mathrm{MDM}\right)\right)},\label{eq:d_covEval}
\end{align}
and $\lambda_i\left(\hbfQ(\bfB),\hbfQ^\mathrm{MDM}\right)$ stands for the roots of the characteristic polynomial $q(\lambda) = \det(\lambda\hbfQ(\bfB)-\hbfQ^\mathrm{MDM})$. 

Note that other metrics based on the higher-order moments of $\bfw_k$ or the estimated density \eqref{eq:pwEstim}  smoothness \citep{DaMaHa:08} can also be used.



\section{Numerical Illustration}
The developed KDD-SNI method is illustrated in two numerical examples including one and two dimensional models and compared with the state-of-the-art methods, that however assume the known state noise PDF family. The introduced numerical experiments demonstrate:
\begin{itemize}
    \item Quality of state noise PDF estimate,
    \item Method robustness w.r.t. the assumed measurement noise model, and
    \item Importance of correct state noise PDF for optimal performance of state estimation and fault detection algorithms.
\end{itemize}

\subsection{Scalar LTV Model}
In the first example\footnote{Example is implemented in \verb|stateNoiseIdent_LTV_1D.m|.}, we consider the scalar model \eqref{eqSt}, \eqref{eqMe} with $F_k=0.9\sin(10^{-4}k), G_k=\cos(k), H_k=1$, $u_k=1$, and $\tau=10^6$. The  measurement noise PDF is zero\discretionary{-}{-}{-}mean Gaussian with unit variance $R=1$, i.e., $p_v(v_k)=\calN\{v_k;0,R\}$. The estimated state noise PDF $p_w(w_k)$ is set for the purpose of simulation as the Rayleigh PDF with the scale-related parameter $\sigma^2=2^2$, i.e., $p_w(w_k)=\calR\{w_k;2\}$. It means that the variance of the state noise is $Q\approx1.72$.

\subsubsection{Noise Identification Performance}
In the first experiment of the example, we compare performance of the proposed \textit{KDD-SNI method} providing the estimate $\hat{p}^\mathrm{KDD}_\bfw(\bfw)$ with the following methods
\begin{enumerate}
    \item \textit{Method of moments} (MoM) assuming known state noise (Rayleigh) PDF family \citep{Ka:93} 
    with estimated parameters. In this case, we can calculate the  parameter ${\sigma}^2$ 
    from the first two non-central state noise moments estimated by the correlation MDM \citep{KoDuSt:18a}. The moments and the parameter are related as
    \begin{subequations}
        \begin{align}
            {\mean}[w]&=\sqrt{{{\sigma}^2}\tfrac{\pi}{2}},\\
            {\mean}[w^2]&=2{{\sigma}^2},
        \end{align}
    \end{subequations}
    and the parameter estimate is computed as
    \begin{align}
        \widehat{\sigma^2}&=\operatorname*{argmin}_{\sigma^2}
            \left\lVert
            \begin{bmatrix}
                \hspace{2mm}\hat{\mean}[w]_{\mathrm{MDM}}-\sqrt{{\sigma}^2\frac{\pi}{2}}\\[1mm]
                \hspace{-4mm}\hat{\mean}[w^2]_{\mathrm{MDM}}-2\sigma^2
            \end{bmatrix}
            \right\rVert^2.
    \end{align}
    The state noise PDF identified by the MoM is denoted as $\hat{p}^{\mathrm{MoM},R}_\bfw(\bfw)$.
    \item MoM assuming the state noise PDF in the form of a ``generic'' Gaussian sum (GS) with $d=1,2,3,4,5$ terms, i.e., MoM designed under the assumption of \textit{unknown} state noise PDF approximated by a GS PDF. The relations between the GS moments and parameters can be found in \citep{Ka:93,DuKoSt:20}. The resulting PDF estimate is denoted as $\hat{p}^{\mathrm{MoM},GS(d)}_\bfw(\bfw)$.
\end{enumerate}

\begin{figure*}
        \hspace*{31.2mm}
        \tikz{\draw[line width=0.3mm] (0,0) -- (0,4.65);}
        \hspace*{126mm}
        \tikz{\draw[color=black, thick, rounded corners, fill = blue!1!white] (0,0) rectangle (2.5,4.65);}
        
        \vspace*{-46.55mm}
        \hspace*{-3mm}
        \includegraphics[width=1.018\textwidth]{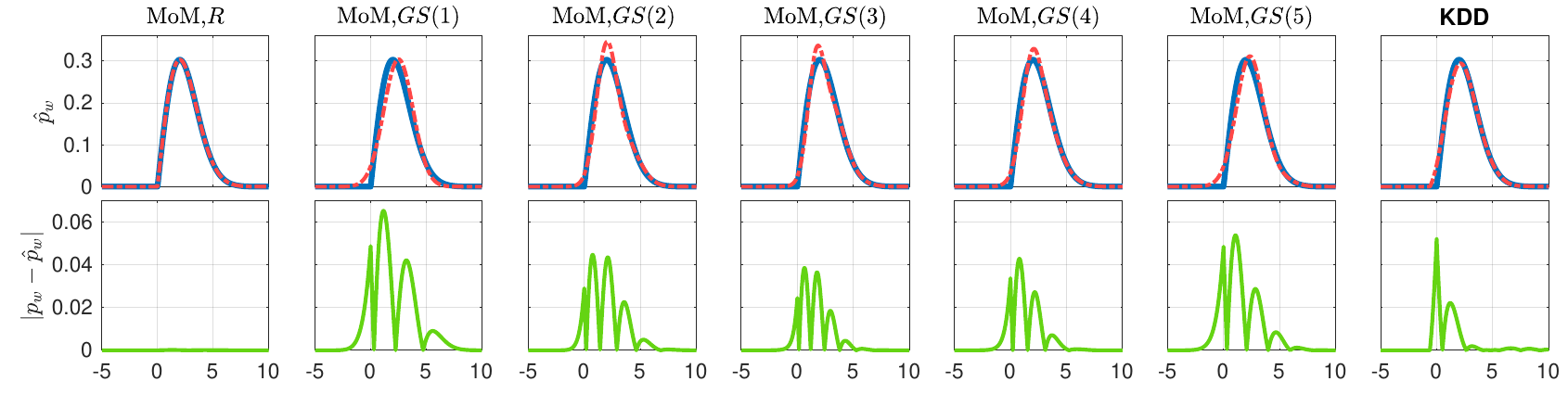}
        
        \vspace*{-1mm}
        \caption{Illustration of true \true \ and estimated state noise PDFs \est, and estimate errors \error.}
        \label{fig:scalarLTV}
\end{figure*}

\begin{figure}
        \includegraphics[width=0.91\columnwidth]{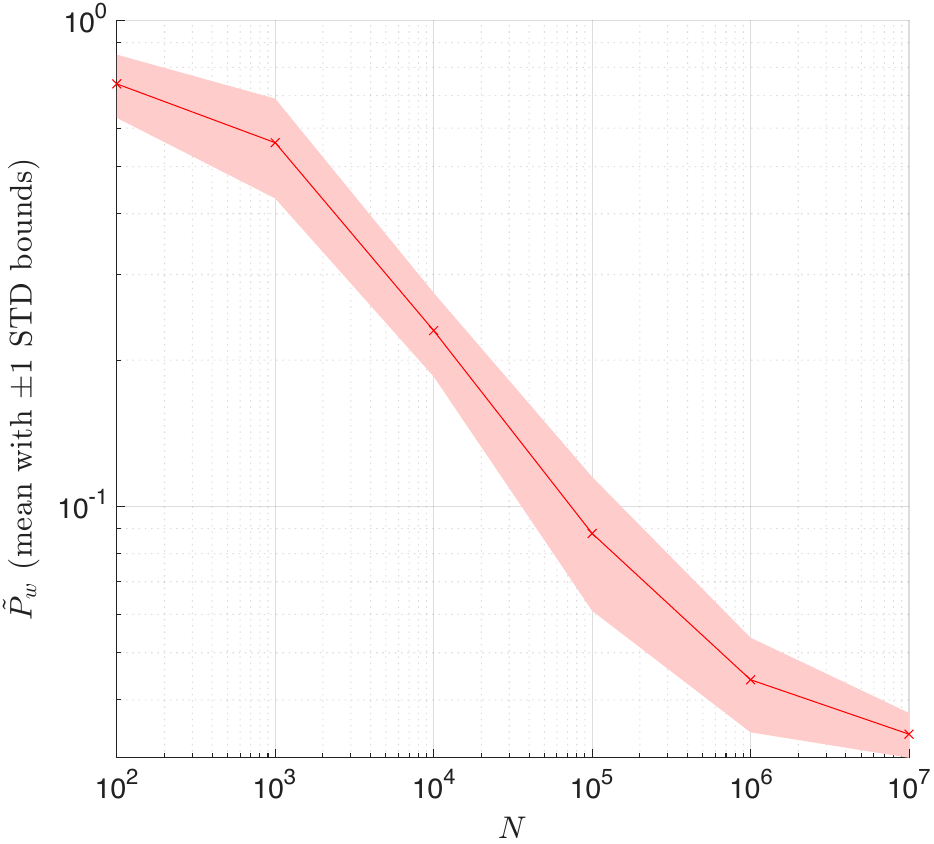}
        \vspace*{-5mm}
        \centering
        \caption{Illustration of convergence by integral error of identified state noise PDF (mean with the uncertainty region $\pm$1 STD).}
        \label{fig:KDDconv}
\end{figure}

The true and estimated state noise PDFs are plotted in Figure \ref{fig:scalarLTV} (top plots) together with the PDF absolute estimation error (bottom plots)
\begin{align}
    \tilde{p}_w(w)=|p_w(w)-\hat{p}_w(w;\bfXi)|.
\end{align}
The integral error of the state noise PDF estimate is defined as
\begin{align}
    \tilde{P}_w=\int \tilde{p}_w(w) dw,\label{eq:ie}
\end{align}
and is summarised in Table \ref{tab:IE_1D}. The results demonstrate that the most accurate method is the MoM based on the assumption of the known correct PDF family (i.e., MoM,$R$) as it could be expected. Relaxing this assumption and \textit{considering} the state noise PDF to be GS distributed leads to significantly worse PDF estimate quality. In these GS cases, we should highlight that increasing the number of the Gaussian terms $d$ does not generally lead to decreased error. The reason lies in the fact that each additional Gaussian term requires estimating three parameters: weight, mean, and variance. This, in turn, necessitates estimating three additional moments of the state noise by the MDM for the MoM. We observe that estimating higher-order moments with the same quality requires more data; in other words, given the fixed sample size, the uncertainty of moment estimates increases with increasing order of estimated moment. Compared to the MoM, the proposed \textit{KDD-SNI method} provides \textit{accurate} estimates \textit{without} any prior knowledge of the state noise PDF.  

Figure \ref{fig:KDDconv} then underlines accurate estimates of the proposed KDD-SNI method and illustrates the convergence of the noise PDF estimate with increasing number of samples $\kappa$ as proven in Section 4. In the figure, the sample mean of the integral error $\tilde{P}_w$ \eqref{eq:ie} calculated over $M=10^3$ Monte-Carlo (MC) simulations is plotted by the red line and the uncertainty bounds defined as $\pm1$ standard deviation (STD) are visualised as light red region along the line. It can be seen that the integral error $\tilde{P}_w$ tends to zero with increasing number of samples $N$ in terms of mean and STD.

\begin{figure}
        \includegraphics[width=1\columnwidth]{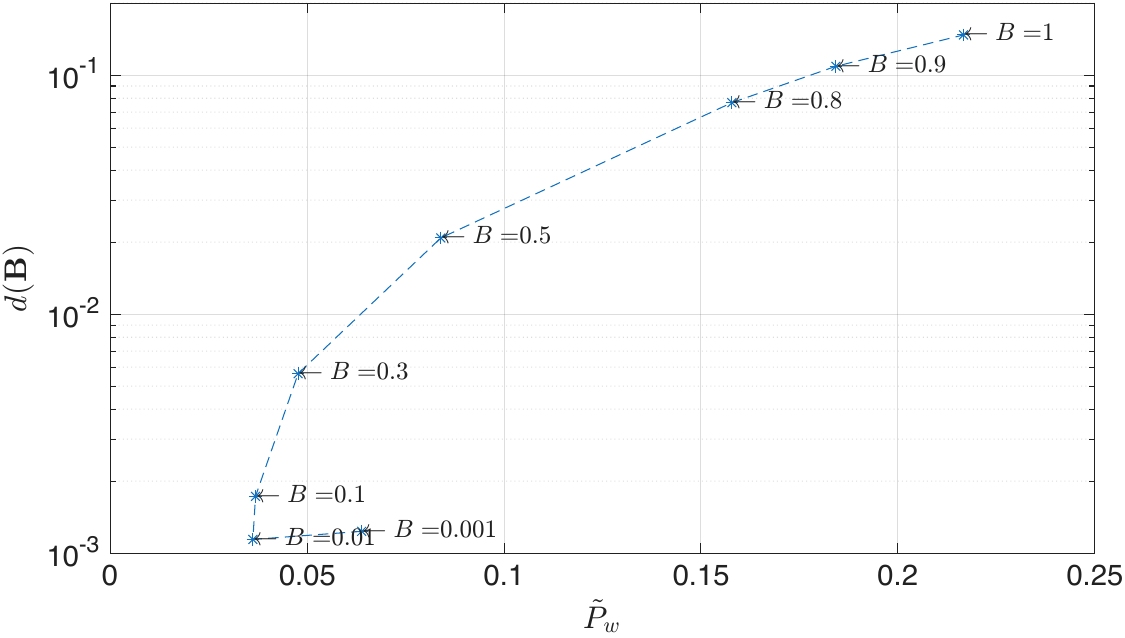}
        \vspace*{-5mm}
        \centering
        \caption{Relation of metric for selection of kernel bandwidth and resulting estimated PDF integral error.}
        \label{fig:criterion_vs_qual}
\end{figure}

The remaining illustration in Figure \ref{fig:criterion_vs_qual} shows relation between the integral error $ \tilde{P}_w$ \eqref{eq:ie} and the value of the metric $d(\bfB)$ \eqref{eq:d_covEval} for different values of the kernel bandwidth $B$. It can be seen that, generally, the metric reliably predicts the error of the identified state noise density (except of the small bandwidth values, which however leads to similar errors).

\begin{table*}
\centering
\begin{tabular}{c|c|ccccc!{\vrule width 0.5mm}c}
                    & MoM,$R$ & MoM,$GS(1)$ & MoM,$GS(2)$ & MoM,$GS(3)$ & MoM,$GS(4)$ & MoM,$GS(5)$ & \textbf{KDD}\\[0.3mm] \hline & & & & & & \\[-3mm]
$\tilde{P}_w$       &  0.0005 & 0.1949 & 0.1193 &  0.0837  & 0.0886  &  0.1398 & 0.0570
\end{tabular}
\vspace*{1mm}
\caption{Integral error of state noise PDF estimates (1D example).}
\label{tab:IE_1D}
\end{table*}



\subsubsection{Method Robustness}
In the second experiment, the robustness of the proposed method w.r.t. accuracy of the measurement noise model is treated. The KDD-SNI method is based on the assumption of the known measurement noise PDF $p_v(v_k)$, which is an acceptable assumption in many cases as the PDF can be found, e.g., by a sensor calibration \citep{Ro:03,Gr:08,KoDuStDa:23}. To illustrate the KDD-SNI method robustness against imprecise knowledge of the measurement noise PDF $p_v(v_k)$, we assume the data are still generated according to the Gaussian PDF with $R=1$, but the KDD-SNI is provided with a different variance value within the range $R^\mathrm{KDD}\in[0.8, 0.85,\ldots,1.2]$. The impact of ``incorrect'' measurement noise variance on the KDD-SNI performance can be found in Table \ref{tab:IE_1D_Rincor} in terms of the integral error \eqref{eq:ie}. It is worth noting that the KDD-SNI is less sensitive to  $R^\mathrm{KDD}<R$ and more sensitive for $R^\mathrm{KDD}>R$, when KDD-SNI trusts that the data variations are due to $w_k$, but in reality a share of the variations is due to $v_k$.

\begin{table*}
\centering
\begin{tabular}{c|cccc|c|cccc}
   $R^\mathrm{KDD}$                 & 0.8& 0.85 & 0.9 & 0.95 & 1 & 1.05 & 1.1 & 1.15 & 1.2\\[0.3mm] \hline & & & & & & & & & \\[-3mm]
$\tilde{P}_w$       & 0.104  & 0.091 & 0.078 & 0.066   & 0.057  & 0.057  & 0.058 & 0.061 & 0.065
\end{tabular}
\vspace*{1mm}
\caption{Integral error of state noise PDF estimates assuming incorrect measurement noise variance $R^\mathrm{KDD}$ (1D example).}
\label{tab:IE_1D_Rincor}
\vspace*{-3mm}
\end{table*}

\subsubsection{Impact on State Estimation and Fault Detection Performance}
The third experiment highlights the importance of the correct state noise PDF description for optimal performance of the state estimation and fault detection. The \emph{first part of this experiment} demonstrates this aspect in the state estimation. In particular, we design:
\begin{itemize}
    \item Kalman filter (KF) \citep{Si:06} based on the identified $p_w(w_k)$ in the form of the approximate Gaussian PDF by the MoM,$GS(1)$, 
    \item Variational Bayesian adaptive KF (VBKF) estimating the state and state noise variance $Q$ under assumption of Gaussian density \citep{MaLaWaZeWaPaMo:18}, and
    \item Grid-based filter (GbF) \citep{Be:99} based on $p_w(w_k)$ identified by (\textit{i}) the  MoM,\textit{GS}(1), (\textit{ii}) the \textit{proposed} KDD.
\end{itemize}
The state noise densities used by the filters, namely densities identified by  the KDD-SNI method and MoM,\textit{GS}(1), are illustrated in Figure \ref{fig:scalarLTV}. The KF can be understood as the \textit{linear minimum mean square error estimator} (LMMSE) with offline-identified state noise moments without the need of any distribution assumption. The VBKF is, on the other hand, a version of the KF with online identification of the state noise variance (the mean is assumed to be known). The GbF is designed with 1000 points enabling very accurate numerical solution to the Bayesian recursive relations. 


The quality of the state estimation is compared in $M$ MC simulations using the following two criteria: 
\begin{itemize}
    \item Root mean square error
    \begin{align}
        \mathrm{RMSE}=\sqrt{\tfrac{1}{M\tau}\sum_{m=1}^M\sum_{k=1}^\tau(x_k-\hat{x}_{k|k}^{(m)})^2},
    \end{align}
    \item Averaged STD
    \begin{align}
        \mathrm{ASTD}=\sqrt{\tfrac{1}{M\tau}\sum_{m=1}^M\sum_{k=1}^\tau P_{k|k}^{(m)}},
    \end{align}
\end{itemize}
where $\hat{x}_{k|k}^{(m)}\approx\mean[x_k|z^k]$ is the filtering estimate of the mean and $P_{k|k}^{(m)}\approx\var[x_k|z^k]$ is the variance provided by the considered filters in $m$-th MC simulation.

Whereas the RMSE assesses the \textit{accuracy} of the filter (the lower, the better), comparison of the RMSE and ASTD assesses the \textit{consistency} (for a consistent estimate, RMSE and ASTD are equal). The results,  summarised in Table \ref{tab:estimPerf}, confirm the utmost importance of the accurate knowledge of the state noise PDF. The KF, VBKF, and even the GbF with the state noise PDF identified by the MoM 
provide less accurate and inconsistent estimates compared to the GbF. The GbF, on the other hand, fully exploits the knowledge of the identified PDF by the proposed KDD-SNI method. Generally, only the full knowledge of the state noise PDF enables the design of the \textit{optimal} (and thus consistent) Bayesian state estimator.

\begin{table}
\centering
\begin{tabular}{l|cccc}
                    & KF & \!VBKF\! & \!GbF(MoM,\textit{GS}(1))\! & \!\textbf{GbF(KDD})\!\\\hline
        RMSE & 0.57 & 1.05 & 0.71 & {0.47} \\
        ASTD & 0.47 & 0.05 & 0.45 & {0.47} 
\end{tabular}
\vspace*{1mm}
\caption{Estimation performance of the KF, VAKF, and GbF with identified noise properties.}
\vspace*{-2mm}
\label{tab:estimPerf}
\end{table}

\begin{figure}
        \includegraphics[width=1\columnwidth]{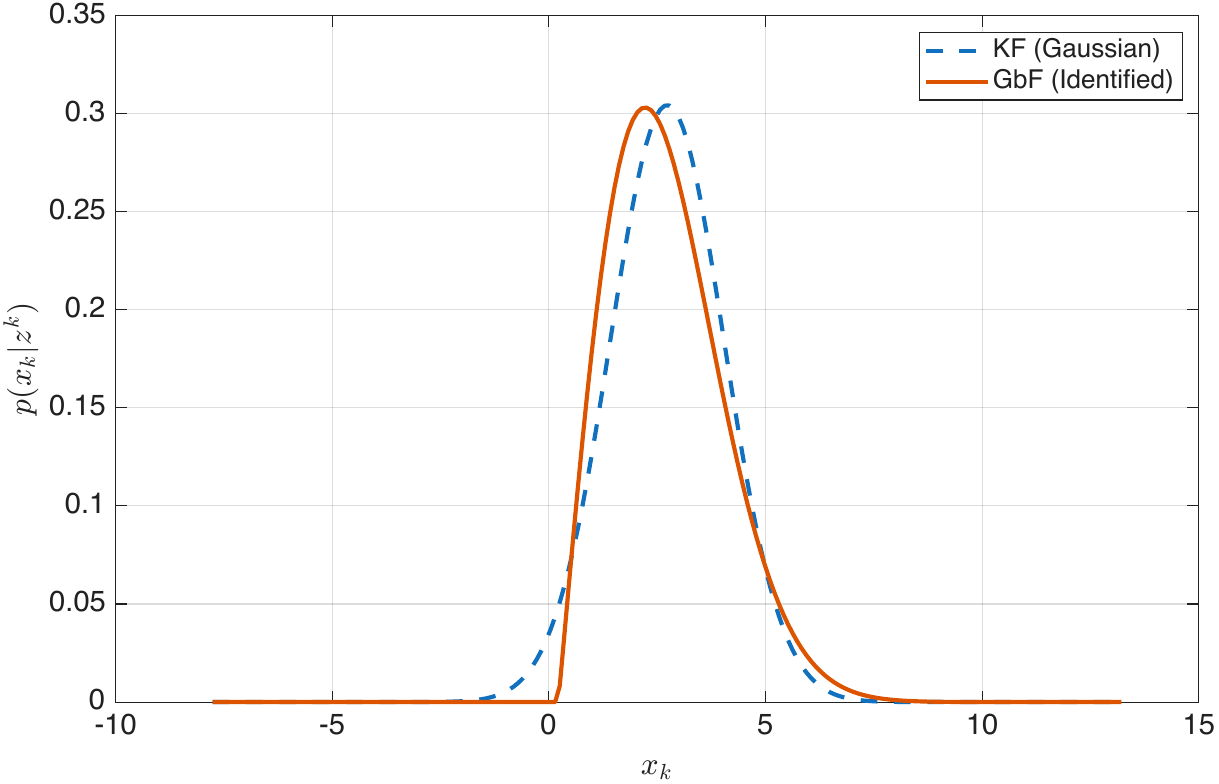}
        \vspace*{-5mm}
        \centering
        \caption{Illustration of conditional PDFs obtained by the (approximate) KF and the (``true'') GbF.}
        \label{fig:condPDF_comp}
\end{figure}

For the sake of completeness, an example of the estimated conditional posterior PDFs are plotted in Figure~\ref{fig:condPDF_comp}. It can be seen that the ``optimal'' PDF provided by the GbF with KDD-identified noise PDF is rather skewed compared to the KF one.


In \emph{the second part of this experiment}, we calculate the quantiles, which are used by fault detection algorithms as a detection threshold\footnote{In the measurement domain, the measurement-predictive PDF is used for a thresholds calculation, by whose comparison w.r.t. the incoming measurement $\bfz_k$ it is decided whether the measurement is correct (i.e., it follows the assumed description) or not \citep{Si:06}.}. The detection threshold can be calculated if and only if the state (or measurement) predictive estimate PDF is known. In case of incorrect PDF, the thresholds may differ leading to, e.g., unreliable detection of faulty measurements.

\begin{table}
\centering
\begin{tabular}{l|cc}
                    & {Lower quant.} & Upper quant.\\\hline
        KF (Gaussian-assumed) & 0.06 & 5.12 \\
        GbF (\textbf{identified})       & 0.58 & 5.53 
\end{tabular}
\vspace*{1mm}
\caption{Lower and upper quantiles of the conditional PDF provided by the KF and the GbF.}
\vspace*{-2mm}
\label{tab:FD}
\end{table}

Let the lower and upper quantiles of the Gaussian\discretionary{-}{-}{-}assumed posterior PDF produced by the KF and the ``optimal'' posterior PDF provided by the GbF be calculated. The lower and upper quantiles are calculated for the probability of false alert $P_\mathrm{FA}=0.95$ allocated equidistantly between them and are summarised in Table \ref{tab:FD}. It can be seen that the quantiles of the state estimate PDF calculated under assumption of the Gaussian state noise PDF $p_\bfw(\bfw_k)$ differ significantly from the true ones based on the identified ``non-Gaussian'' $p_\bfw(\bfw_k)$. Such incorrectly specified quantiles would lead to non-optimal performance of a fault detection algorithm.

\subsection{Two-dimensional Model: UWB-based Navigation}
\begin{table}
\centering
\begin{tabular}{c|ccc}
                    & MoM,$GS(1)$ & MoM,$GS(2)$ & \textbf{KDD}\\[0.3mm] \hline & & & \\[-3mm]
$\tilde{P}_w$       &  0.392 &  0.040   &  0.198 
\end{tabular}
\vspace*{1mm}
\caption{Integral error of state noise PDF estimates (2D example).}
\label{tab:IE_2D}
\end{table}

In the second example\footnote{Example is implemented in  \verb|stateNoiseIdent_LTV_nx2_nz3_tracking|.}, we consider the ultra-wideband navigation in horizontal plane with three anchors. In particular, the \textit{state} $\bfx_k$ with $n_x=2$ represent the vehicle horizontal position and its \textit{dynamics} \eqref{eqSt} is modelled by the random walk model, where the matrices are defined as
\begin{align}
    \bfF_k=\bfI_2,\ \bfG_k=\delta t\bfI_2,
\end{align}
$\delta t=0.1 [s]$ is the sampling period, $\bfu_k$ is the velocity\footnote{For realistic data generation, the velocity is generated using the Ackerman model with varying speed and sinusoidal steering.} measurement, $\bfI_2$ stands for the identity matrix of indicated dimension, and the state noise PDF (used for data generation \textit{only}) is the two-component Gaussian sum PDF $p_\bfw(\bfw_k)=\tfrac{1}{2}\sum_{j=1}^2\{\bfw_k;\bar{\bfw}^{(j)},\bfQ^{(j)}\}$, where 
\begin{align}
    \bbfw^{(1)}&=\left[\begin{smallmatrix}
        -2\\ \ \ 2
    \end{smallmatrix}\right],
    \bbfw^{(2)}=\left[\begin{smallmatrix}
        \ \ 2\\ -2
    \end{smallmatrix}\right],\nonumber\\
    \bfQ^{(1)}&=\left[\begin{smallmatrix}
        4& 1.5\\ 1.5& 2
    \end{smallmatrix}\right], 
    \bfQ^{(2)}=\left[\begin{smallmatrix}
        1& 0.5\\ 0.5& 2
    \end{smallmatrix}\right].
\end{align}
The covariance matrix of the state noise is $\bfQ\approx\left[\begin{smallmatrix}
        6.5& -3\\ -3& \ \ 6
    \end{smallmatrix}\right]$.

The \textit{measurement} $\bfz_k$ with $n_z=3$ represents the  distance between the vehicle and the $i$-th anchor with the known location $\mathbf{l}_i$, and the linearised measurement equation is defined as
\begin{align}
    \bfz_{k,i}\approx\bfh_{k,i}(\hbfx_{k}) +\bfH_{k,i}(\bfx_k-\hbfx_{k})+\bfv_k,
\end{align}
where $\bfh_{k,i}(\bfx_k)=||\mathbf{l}_i-\bfx_k||_2$, $||\cdot||_2$ denotes the Euclidean norm, $\bfH_{k,i}=\tfrac{\partial\bfh_{k,i}(\bfx_k)}{\partial\bfx_kk}|_{\bfx_k=\hbfx_k}$ is line-of-sight (LoS) vector forming the measurement matrix $\bfH_k$ in \eqref{eqMe}, and $\hbfx_k$ is a linearisation point. 
The measurement noise is described by the Gaussian PDF $p_\bfv(\bfv_k)=\calN\{\bfv_k;\bfI_{3\times1},\bfR\}$ with $\bfR=\left[\begin{smallmatrix}
    1 &-0.5& 0.2\\ -0.5 &1 &0.3\\ 0.2 &0.3 &1
\end{smallmatrix}\right]$. 

\begin{figure*}
        \includegraphics[width=0.95\textwidth]{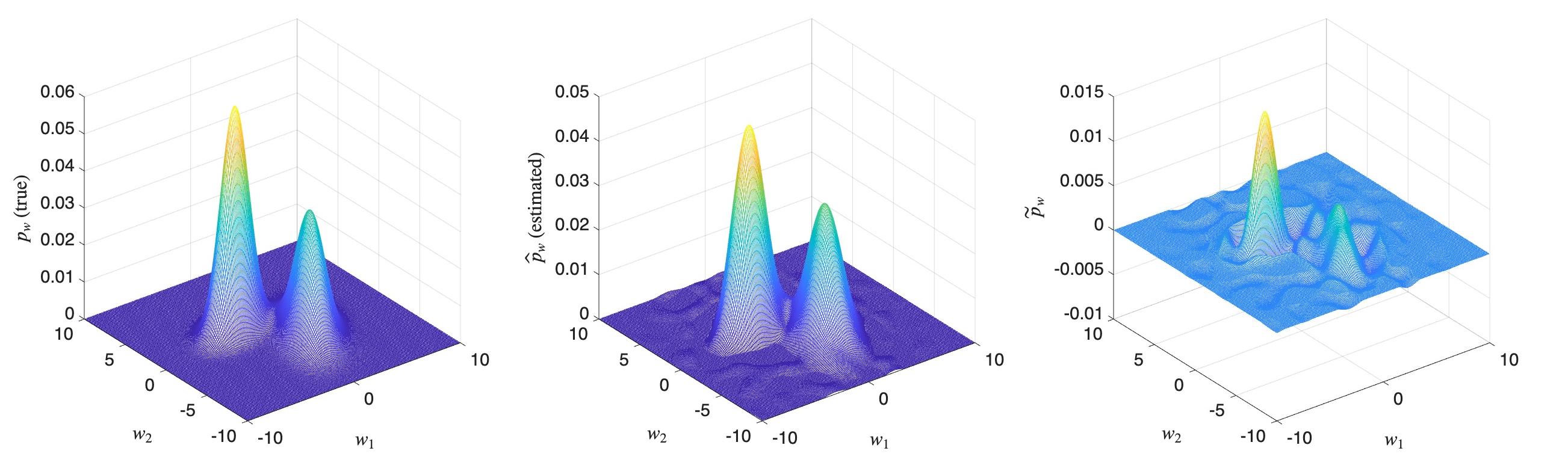}
        \vspace*{-5mm}
        \caption{Illustration of state noise PDFs: \textit{(i)} true $p_\bfw(\bfw_k)$, \textit{(ii)} KDD-SNI estimated $\hat{p}_\bfw(\bfw_k;\bfXi)$, and \textit{(iii)} estimate error $\tilde{p}_\bfw(\bfw_k)$; all evaluated at grid $\bfXi$.}
        \label{fig:vectorLTI_pw}
\end{figure*}

In this example, we focus on the first and third experiments. In the first experiment, we compare the accuracy of the state noise PDF identified by the KDD-SNI method and the MoM. The results can be seen in Table \ref{tab:IE_2D}. Analogously to the results of the previous example, the MoM,\textit{GS}(2), which in fact, assumes the known PDF family, leads to the most accurate estimates. The assumption of a single Gaussian, i.e., MoM,\textit{GS}(1), leads to the significantly worse PDF estimate. On the other hand, the proposed KDD-SNI method results in moderate error, but \textit{without any} assumption on the distribution family. 
The true and KDD-SNI estimated state noise PDFs are given in Figure~\ref{fig:vectorLTI_pw} together with the estimate error. 

\begin{figure}
        \includegraphics[width=\columnwidth]{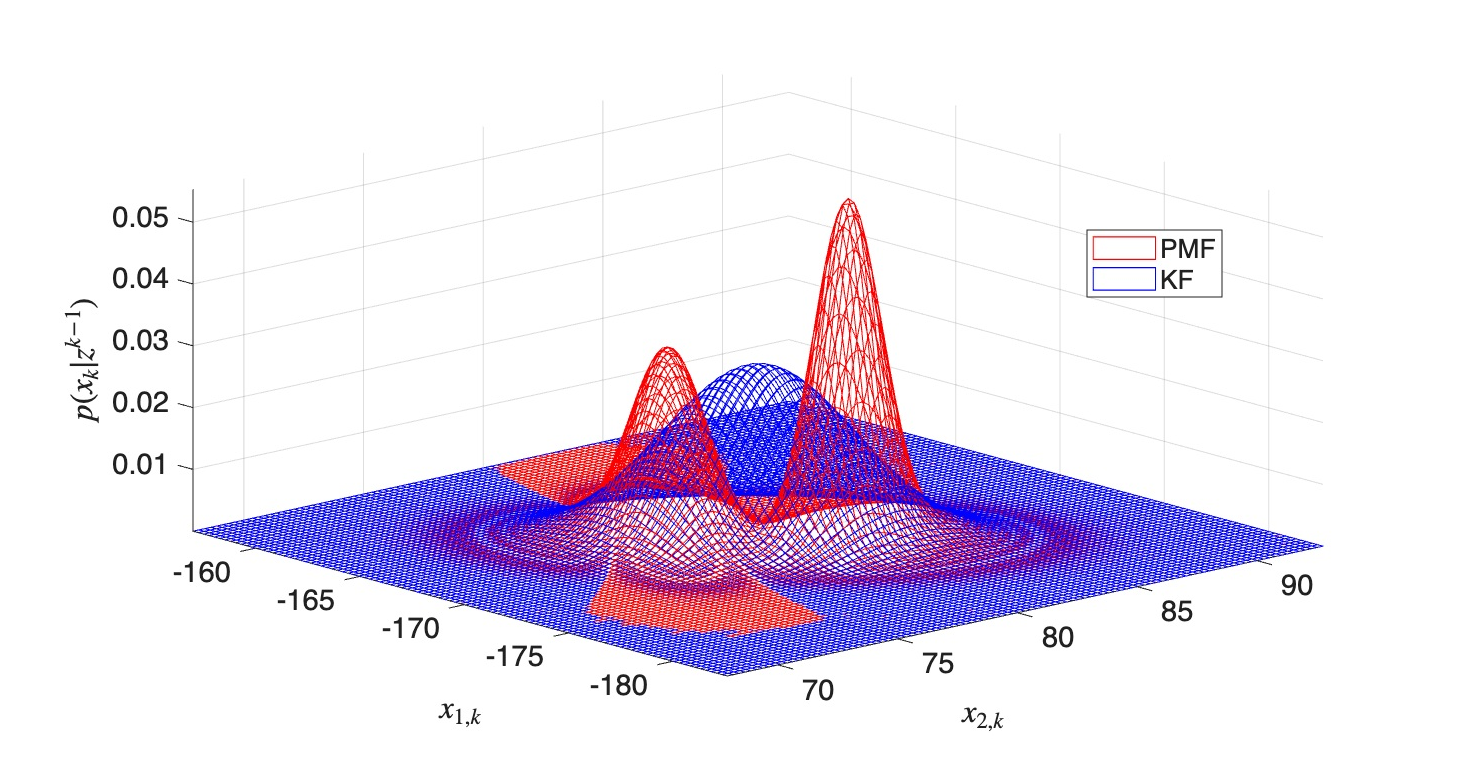}
        \vspace*{-5mm}
        \caption{Illustration of conditional PDFs obtained by the Gaussian-assumed KF and the PMF.}
        \label{fig:pdf_PMF_KF}
\end{figure}

For the sake of completeness, in the next experiment the predictive PDFs obtained by the KF (based on the assumption of Gaussian state noise PDF) and the PMF (based on the identified multimodal PDF) are compared and shown in Figure \ref{fig:pdf_PMF_KF}, where the impact of correct noise modelling is clearly seen.


\subsection{Computational Complexity}
\begin{figure}
        \includegraphics[width=\columnwidth]{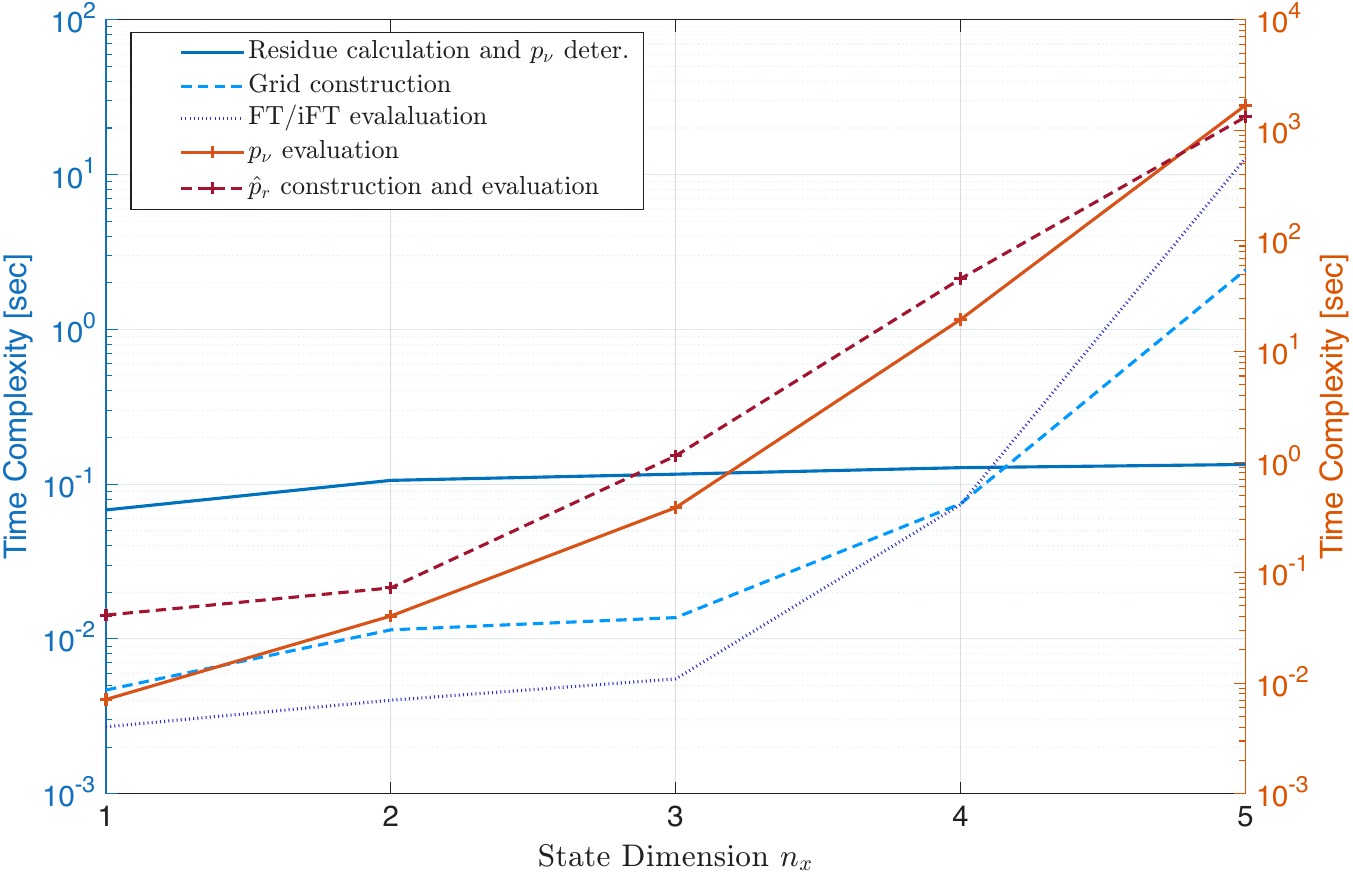}
        \vspace*{-5mm}
        \caption{Illustration of KDD-SNI components computational complexity scaling with dimension.}
        \label{fig:compcompl}
\end{figure}

The proposed algorithm summarised in Section \ref{sec:alg} relies on the following main operations,  Steps (iii)-(x), with computational complexity depending on the state dimension $n_x$. The computational complexity as a function of dimension is shown in Figure \ref{fig:compcompl} for an LTV model with fixed number of grid points per dimension. We can see that the most demanding parts are the construction of the kernel density $\hat{p}_{\bfr_k}$ and evaluation of the densities $p_{\bfnu_k}$, $\hat{p}_{\bfr_k}$ in the grid in \eqref{eq:PrPv_ltv} $\forall k$. To reduce the computational complexity we can employ various algorithmic optimizations and software packages used to calculate continuous probability distributions from discrete data points efficiently. Instead of naive calculations, fast methods drastically reduce computation time for large, multidimensional datasets.

\section{Concluding Remarks}
This paper dealt with the identification of the state noise PDF of a linear system described by the state-space model. The proposed KDD-SNI method uses a specially designed calculation of the residue (or the measurement prediction error), which can be interpreted as the sum of the state noise and linearly transformed measurement noise. Assuming a known measurement noise PDF, calculating the residue, and constructing the respective residue kernel density, we can estimate the state noise PDF by the developed density deconvolution algorithm. 
The derivation is based on random finite set theory. The algorithm can be efficiently realised by the Fourier transform and equipped with a method for automatic selection of user-defined kernel parameters. Moreover, the prosed method provides \textit{consistent} estimate of the state noise PDF \textit{without} any assumption on the knowledge of the PDF family, which is a unique property among the state-of-the-art methods. The proposed algorithm was numerically verified and the paper is supplemented with a code of the proposed method.

\appendix
\section{Time-varying Density Deconvolution from Fusion Perspective}
In this appendix, we provide interpretation of the estimated state noise PDF $\hat{p}_\bfw(\bfw)$ derived from \eqref{eq:Fwestim_ltv_full_concept} from the PDF fusion perspective.
Consider $\kappa$ ``single-sample-based'' estimates $\hat{p}^{(k)}_\bfw(\bfw)$ of the true PDF ${p}^{(k)}_\bfw(\bfw)(\approx{p}_\bfw(\bfw), \forall k)$, each based on the FT $\hat{\calF}_{\bfr_k}(\bft)$ of the KD estimate $\hat{p}_{\bfr_k}(\bfr)$ \eqref{eq:kernel_r_LTV}. Then, we can write
\begin{align}
    {\calF}_{\bfw}^{(k)}(\bft)=\frac{{\calF}_{\bfr_k}(\bft)}{\calF_{\bfnu_k}(\bft)}, \forall k\in \calK,\label{eq:Fwestim2}
\end{align}
where ${\calF}_{\bfr_k}(\bft)$ is the CF of ${p}_{\bfr_k}(\bfr)$ \eqref{eq:kernel_r_LTV}. To fuse \eqref{eq:Fwestim2} using arithmetic mean to a state noise PDF estimate, we can use the \textit{density arithmetic mean} commonly used for the PDF fusion \citep{KoElDjHl:22}. Assuming the same weight $\sfrac{1}{\kappa}$ of each calculated CF, the CF arithmetic mean reads
\begin{align}
    {\calF}_{\bfw}(\bft)=\frac{1}{\kappa}\sum_{k\in\calK}{\calF}_{\bfw}^{(k)}(\bft),\label{eq:Fwestim_all}
\end{align}
which can be further treated, w.r.t. \eqref{eq:Fwestim2}, as
\begin{align}
    {\calF}_{\bfw}(\bft)=\frac{1}{\kappa}\sum_{k\in\calK}\frac{{\calF}_{\bfr_k}(\bft)}{\calF_{\bfnu_k}(\bft)}.\label{eq:Fwestim_all2}
\end{align}
Considering the fact that
\begin{align}
    \left({\calF}_{\bfw}(\bft)=\right)\frac{{\calF}_{\bfr_k}(\bft)}{\calF_{\bfnu_k}(\bft)}=\frac{{\calF}_{\bfr_{k+1}}(\bft)}{\calF_{\bfnu_k+1}(\bft)}, \forall k,
\end{align}
the averaged (or fused) CF \eqref{eq:Fwestim_all2} estimate becomes
\begin{align}
    {\calF}_{\bfw}(\bft)=\frac{\tfrac{1}{\kappa}\sum_{k\in\calK}{\calF}_{\bfr_k}(\bft)}{\tfrac{1}{\kappa}\sum_{k\in\calK}\calF_{\bfnu_k}(\bft)}.\label{eq:Fwestim_all3}
\end{align}
Substitution of the CF $\hat{\calF}_{\bfr_k}(\bft)$ of the KD estimate $\hat{p}_{\bfr_k}(\bfr)$ \eqref{eq:kernel_r_LTV} instead of ${\calF}_{\bfr_k}(\bft)$ leads to Eq. \eqref{eq:Fwestim_ltv_full_concept}.

\bibliographystyle{elsarticle-num}
\bibliography{literatura}
\end{document}